\documentclass[galaxies,article,accept,moreauthors,pdftex,10pt,a4paper]{mdpi}

\firstpage{1}
\articlenumber{x}
\doinum{10.3390/------}
\pubvolume{xx}
\pubyear{2016}
\copyrightyear{2016}
\externaleditor{Academic Editors: Stefano Ettori and Dominique Eckert}
\history{Received: 11 October 2016; Accepted: 23 November 2016; Published: date}
\usepackage{textcomp}
\usepackage{graphicx}
\usepackage{float}
\usepackage{amsmath}
\usepackage{amssymb}
\usepackage{ulem}
\usepackage{array}
\usepackage{enumerate}
\usepackage[usenames]{xcolor}
\usepackage{natbib,twoopt}
\usepackage{comment}

 \theoremstyle{mdpi}
 \newcounter{thm}
 \newcounter{ex}
 \newcounter{re}
\newcommand{\mach}{{\cal M}}

\newcommand{\plk}{{\it Planck~}}
\newcommand{\chan}{{\it Chandra~}}

\newcommand{\amin}{{^{\prime}}}
\newcommand{\asec}{{^{\prime\prime}}}

\Title{Galaxy cluster outskirts from the thermal SZ and non-thermal synchrotron link}

\Author{Kaustuv Basu $^{1,*}$, Jens Erler$^{1}$, Martin Sommer$^{1}$,  Franco Vazza$^{2}$, and Dominique Eckert$^{3}$}
\AuthorNames{Firstname Lastname, Firstname Lastname and Firstname Lastname}

\address{%
$^{1}$ \quad Argelander Institut f\"ur Astronomie, Universit\"at Bonn, Auf dem H\"ugel 71, 53121 Bonn, Germany \\
$^{2}$ \quad Hamburger Sternwarte, Gojenbergsweg 112, 21029 Hamburg, Germany \\
$^{3}$ \quad Department of Astronomy, University of Geneva, Chemin d'Ecogia 16, 1290 Versoix, Switzerland}

\corres{Correspondence: kbasu@astro.uni-bonn.de}

\abstract{Galaxy cluster merger shocks are the main agent for the thermalization of the intracluster medium and the energization of cosmic ray particles in it. Shock propagation changes the state of the tenuous intracluster plasma, and the corresponding signal variations are measurable with the current generation of X-ray and Sunyaev--Zel'dovich (SZ) effect instruments. {Additionally}, non-thermal electrons (re-)energized by the shocks {sometimes} give rise to  extended and luminous synchrotron sources known as radio relics, which are prominent indicators of shocks propagating roughly in the plane of the sky.
In this short review, we discuss how the joint modeling of the non-thermal and thermal signal variations across radio relic shock fronts is helping to advance our knowledge of the gas thermodynamical properties and magnetic field strengths in the cluster outskirts. We {describe the first use of the SZ effect to measure} the Mach numbers of relic shocks, for~both the nearest (Coma) and the farthest (El Gordo) clusters with known radio relics.
}

\conferencetitle{2016 European Week of Astronomy and Space Science, Athens, Greece}

\begin{document}

%%%%%%%%%%%%%%%%%%%%%%%%%%%%%%%%%%%%%%%%%%
%% Only for the journal Gels: Please place the Experimental Section after the Conclusions

%%%%%%%%%%%%%%%%%%%%%%%%%%%%%%%%%%%%%%%%%%

\section{Introduction: cluster outskirts as revealed by shocks}

Despite the oft-quoted phrase of galaxy clusters being the largest virialized objects in the universe, at no stage of their evolution are galaxy clusters truly isolated systems.
In addition to a continuous {matter accretion taking place at the outer boundaries}, clusters grow in mass through dramatic merger events which generate low Mach-number shocks in the hot intracluster medium (ICM). These shocks, {with typical Mach number values in the range 2--4},  produce {sharp jumps} in the density and temperature fields of the ICM, and also energize a population of GeV-energy electrons through the first-order Fermi acceleration process (e.g., \cite{Mini01,Pfrom06,Brug12}). In the rare cases when the merger happens roughly in the plane of the sky, the surface brightness discontinuities become easy to observe and to model. Radio synchrotron emission from cluster shock fronts {viewed roughly edge-on} are known as radio relics, and are typically observed in the cluster outskirts (e.g., review by \cite{Fer12}). Therefore, performing follow-up observations of  radio relics in several other wavebands {and carrying out a comprehensive shock modeling} can be a useful method to learn about the physical conditions in the galaxy cluster outskirts.

The connection between merger shocks and the galaxy cluster radio relics has been established through numerous theoretical works (e.g.,  \cite{HB07,Nuz12,Kang12,Skill13,Vaz14}). Until recently, the observational evidence followed mainly from X-ray data (e.g., \cite{Fino10,Maca11,Aka13,Ogr13}), but now powerful millimeter/\linebreak sub-millimeter instruments like {the MUSTANG/GBT, NIKA/IRAM, and in particular, ALMA
~are ready to provide} resolved images of galaxy cluster merger shocks through the Sunyaev--Zel'dovich (SZ) effect  {(see \cite{Korn11} for the first SZ shock result in MACS J0744.8$+$3927)}.
We have made the first SZ shock measurements for radio relics in the cluster outskirts:  %These measurements were made
for both the nearest radio relic cluster (Coma) using \plk data \cite{Er15}, and also for the most distant radio relic cluster (ACT-CL J0102$-$4915, \linebreak or ``El Gordo'') using ALMA data  \cite{ALMA}. In this {conference proceeding} article, we briefly review these results to illustrate how the joint modeling of the thermal (SZ effect and X-ray) and the non-thermal (radio synchrotron) signals can provide a consistent picture of the shock Mach number and the physical conditions of the ICM in the cluster outskirts. First, however, we start by {describing} how the SZ effect can potentially influence the relic flux measurements at GHz frequencies (1--30 GHz), giving the impression of a rapidly steepening synchrotron spectrum \cite{SZrelic}.

%%%%%%%%%%%%%%%%%%%%%%%%%%%%%%%%%%%%%%%%%%

\begin{figure}[t]
%\centering
\hspace{-6mm}
\includegraphics[width=8cm]{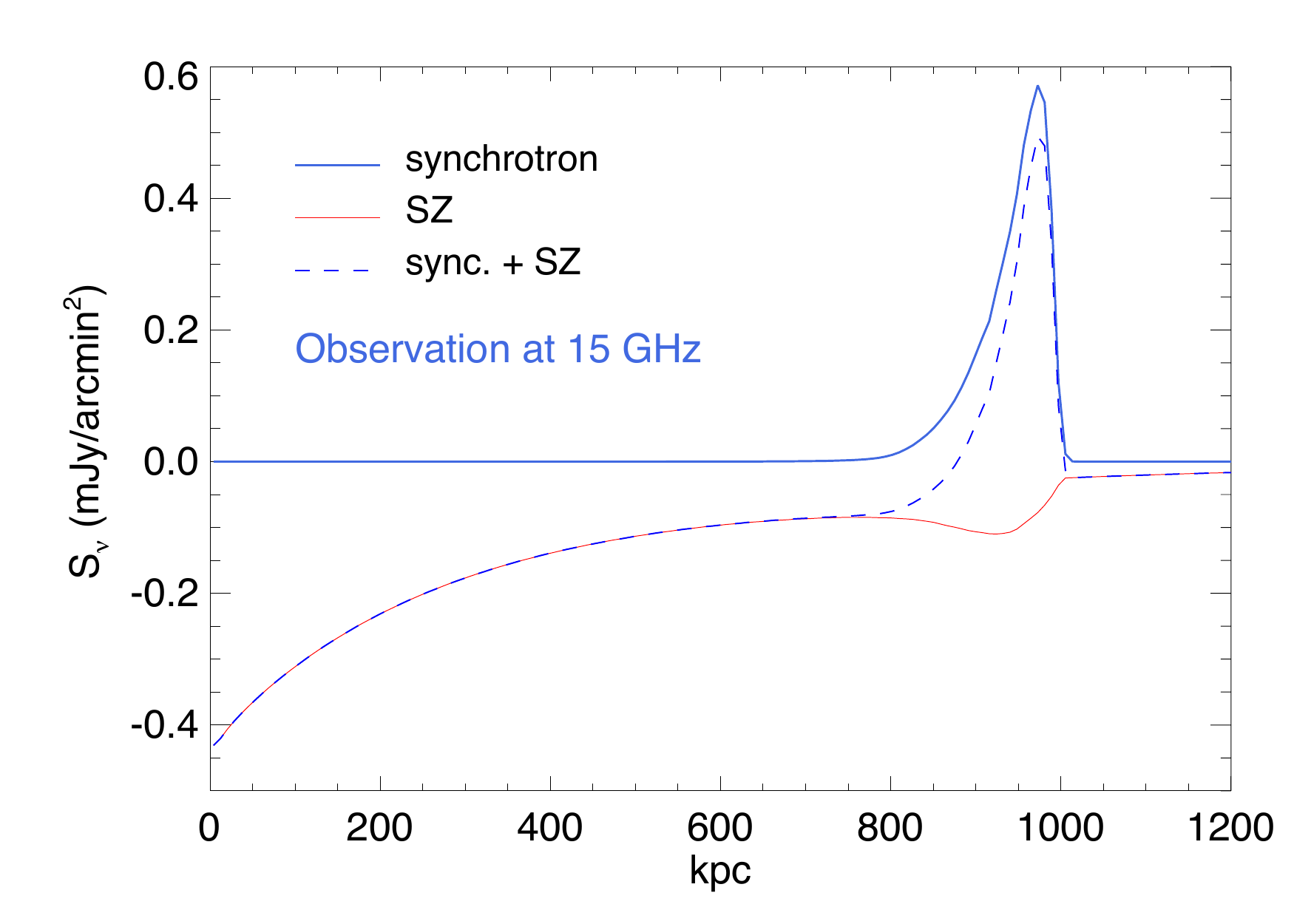}%
\hspace{5mm}
\includegraphics[width=8cm]{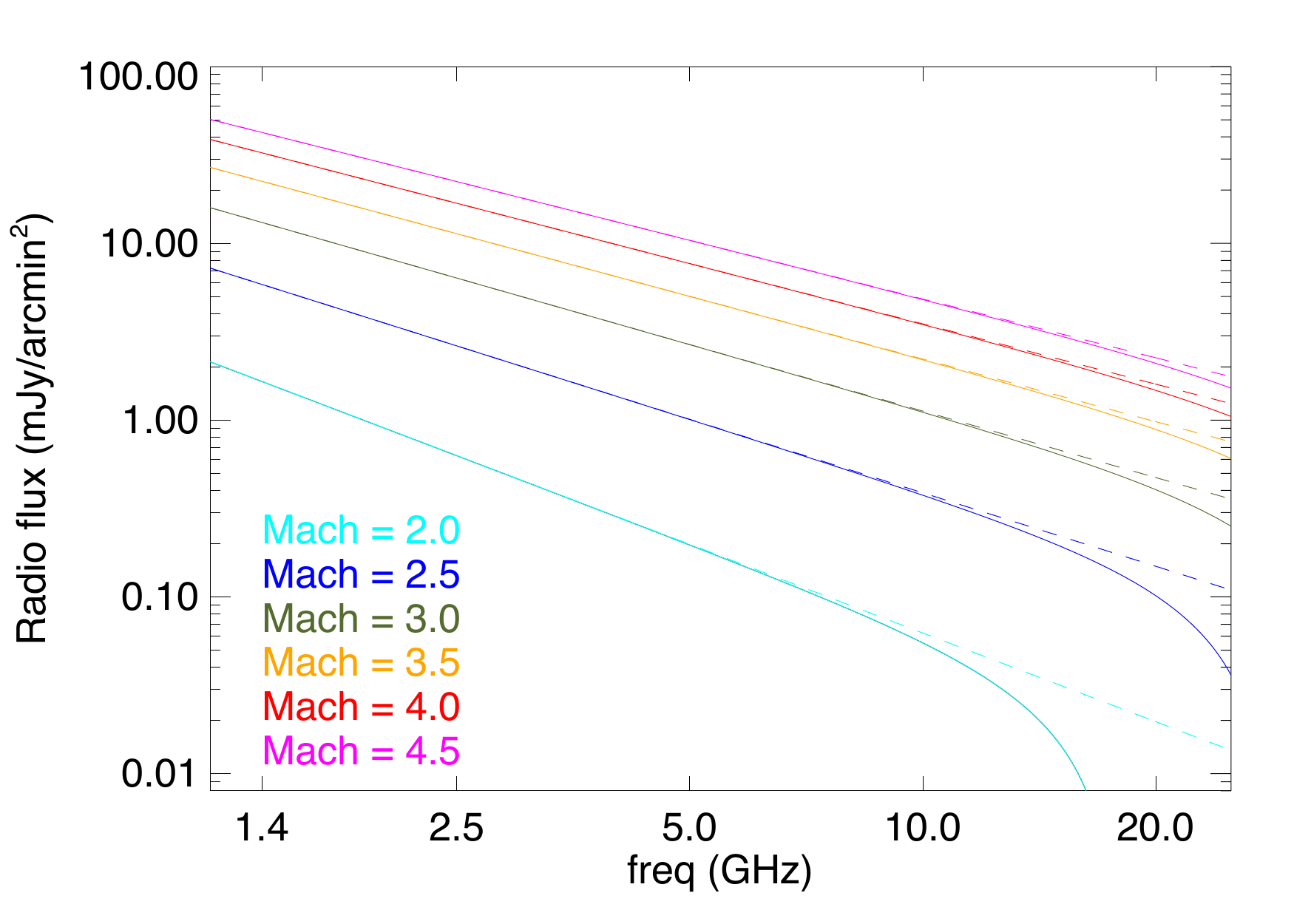}
\caption{Sunyaev--Zel'dovich (SZ) effect contamination {at GHz-frequency observations} of radio relics (from \cite{SZrelic}). \textbf{Left:} Projected synchrotron and SZ effect signals {at the location of a radio relic based on a simple spherical} shock model, illustrating the decrease in the observed synchrotron flux.
%The observed synchrotron flux is less than the true value at 15 GHz due to SZ flux contamination.
\textbf{Right:} Model spectra for radio relics in the GHz frequency range, {based on} a linear Mach dependence of the synchrotron-to-SZ flux ratio. The dashed lines are the uncontaminated spectra, {which follow} a power-law at all frequencies. }
\label{fig:radiosz}
\end{figure}

\section{Relic shocks as seen in radio synchrotron and the SZ effect}
\label{sec:szcont}

A direct consequence of having a shock for producing a radio relic is that there should be an associated boost in the ambient thermal pressure of the ICM, just like the boost in the ICM density and temperature that has been observed in the X-ray.
This boost would then be visible in the SZ effect, which measures the line of sight integral of the gas thermal pressure.
Since the pressure boost roughly scales as Mach number squared, one can easily produce an order of magnitude increase for a $\mach \sim 3$ shock. Thus, even though the low ambient pressure in the cluster outskirts is currently difficult to measure from the SZ effect, this shock-related boost should be easily observable. Moreover, at GHz frequencies, the steeply falling synchrotron spectrum will lead to a situation where the {negative shock-boosted SZ signal amplitude} will become comparable to the {positive} synchrotron flux. A~steepening of the relic spectrum has recently been observed in the famous ``Sausage'' relic in the CIZA J2248.8$+$5301 cluster, and the ``Toothbrush'' relic in 1RXS J0603.3$+$4214 \cite{Str14,Str15}. This has resulted in various theoretical speculations attempting to explain the steepening, which are based on  non-standard extensions of the diffusive shock acceleration (DSA) paradigm \cite{Ka16,Fuji16,Don16}. Our proposed explanation~\cite{SZrelic} is a simple consequence of the relic shock hypothesis, and while the SZ effect may not be the only contributor to the observed spectral steepening, it can be the dominant effect and should be easy to account for.

\begin{figure}[h]
%\centering
\hspace{-6mm}
\includegraphics[height=4.5cm]{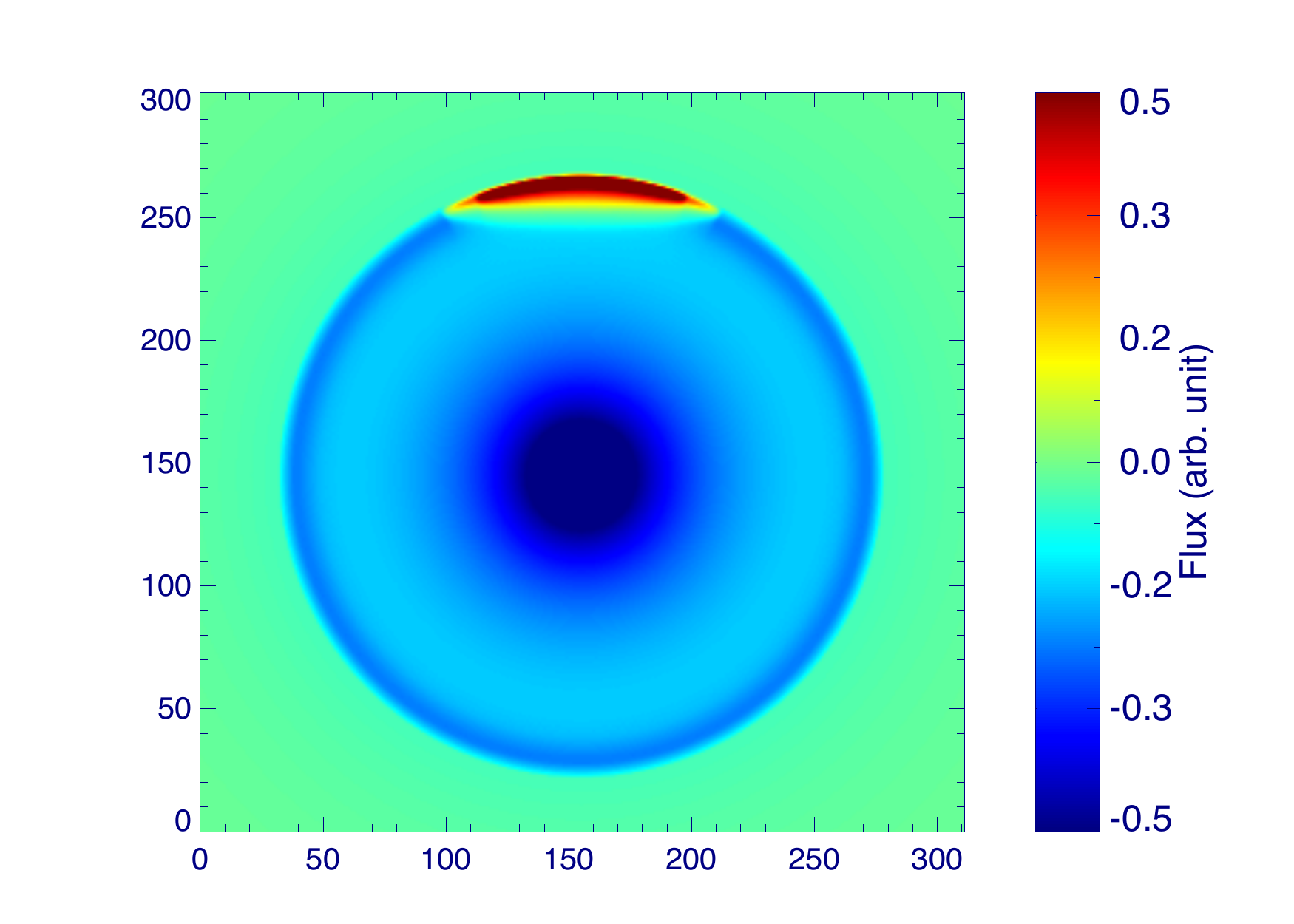}%
\includegraphics[height=4.5cm]{vla_simulations.pdf}
\caption{Simulation of the SZ flux contamination for radio relic observations (from \cite{SZrelic}). \textbf{Left:} A 2D toy model of the synchrotron and SZ flux distribution in a cluster with a spherical shock front. The SZ signal is shown azimuthally extended beyond the radio synchrotron emission region purely for illustration. \textbf{Right:} {A mock but realistic interferometric observation of this radio relic at 10~GHz} with and without the SZ effect. The low relic flux measurement {in the presence of} the SZ signal is~evident.} 
\label{fig:simobs}
\end{figure}   

\begin{figure}[h]
%\centering
\hspace{-6mm}
\includegraphics[width=8cm]{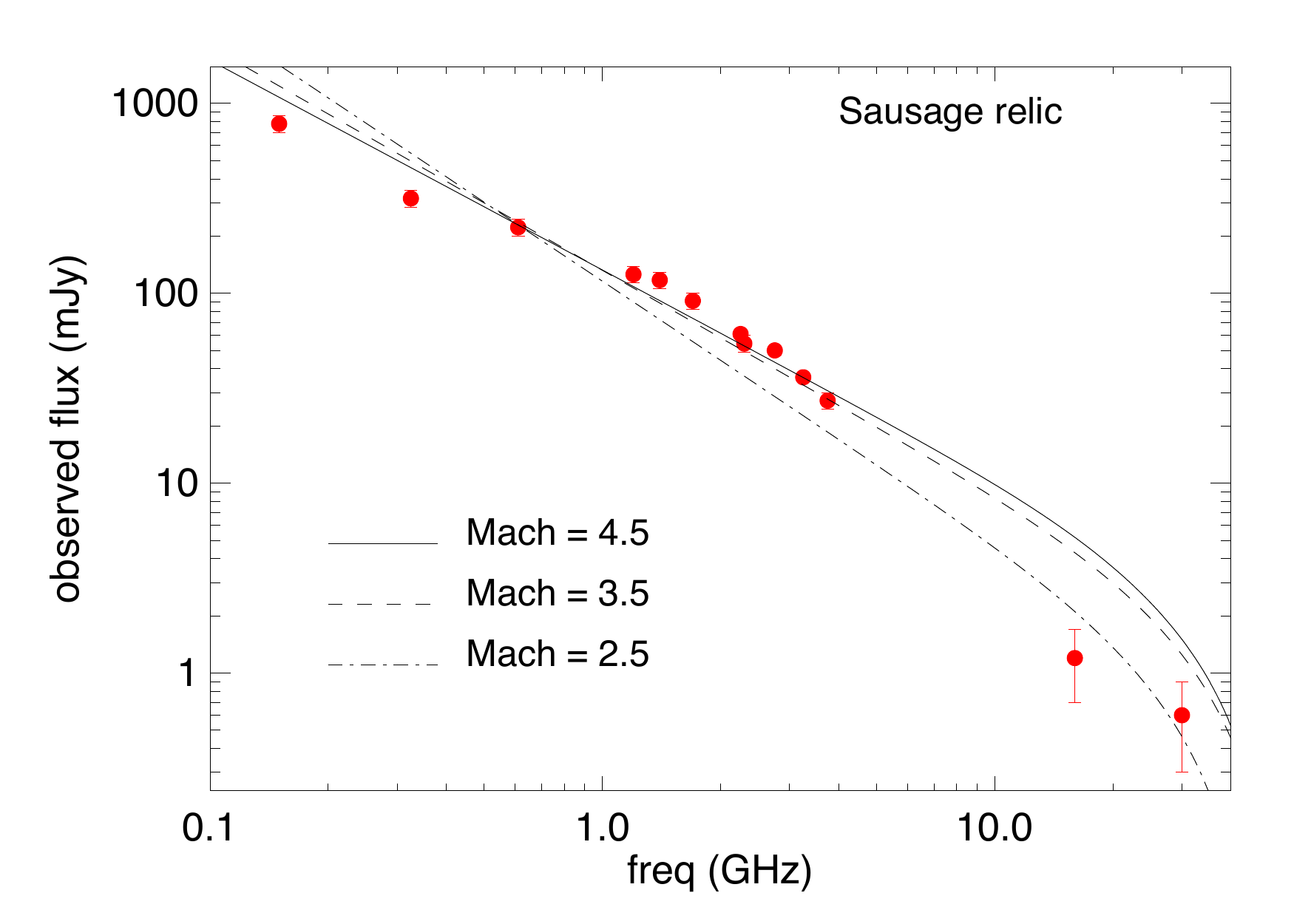}%
\hspace{5mm}
\includegraphics[width=8cm]{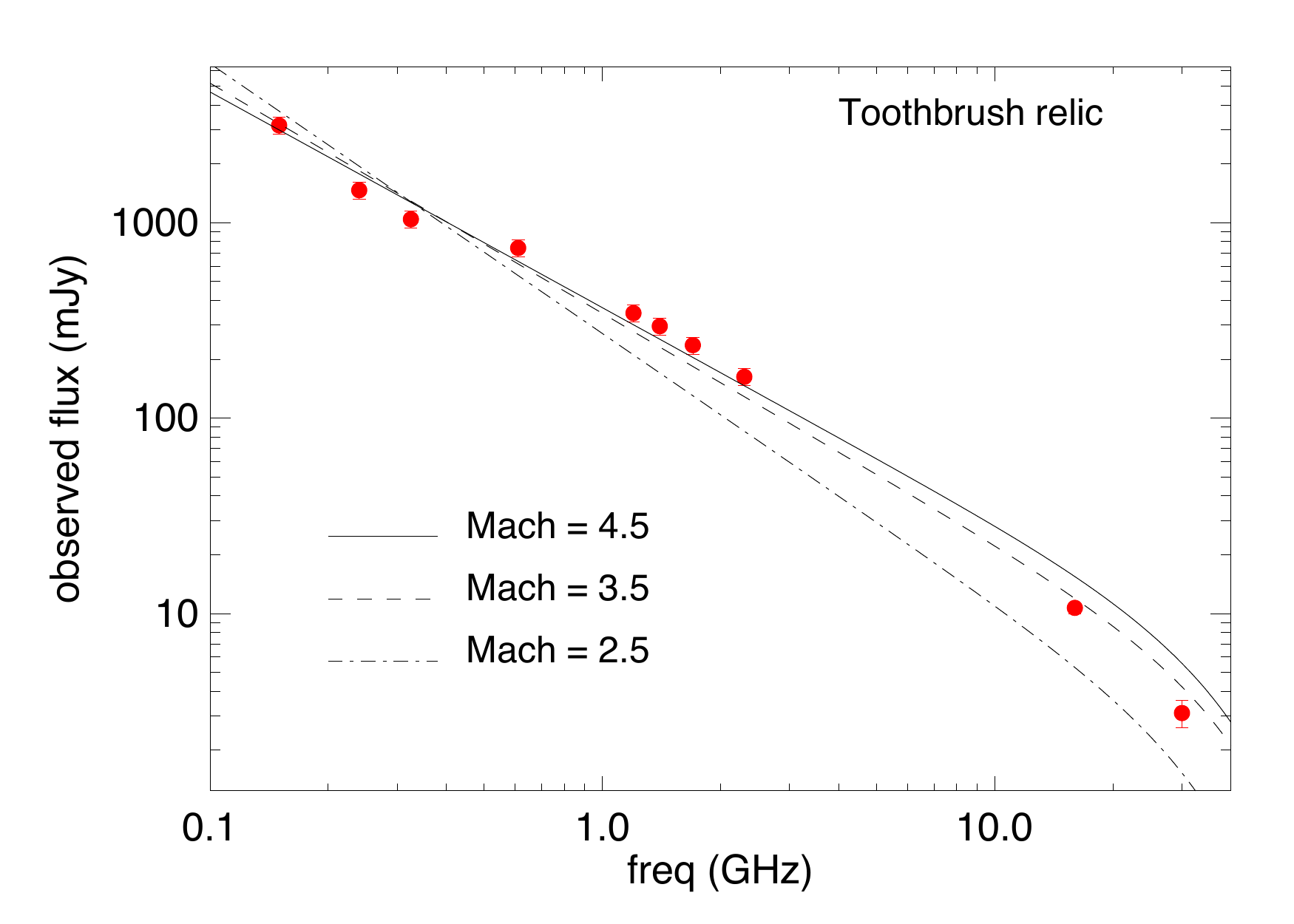}
\caption{Fit to the actual relic flux measurements for the ``Sausage'' relic in CIZA J2248.8$+$5301 (\textbf{left}), and the ``Toothbrush'' relic in 1RXS J0603.3$+$4214 (\textbf{right}), against our model predictions for three different shock Mach numbers  (from \cite{SZrelic}). These model spectra correspond to an ideal total power measurement, and the results from {specific} interferometric imaging can differ.}
\label{fig:relicfit}
\end{figure}

In Figure~\ref{fig:radiosz} (left), we show a simplified model of how {at GHz frequencies} the radio synchrotron and the SZ effect signals may look at the position of a radio relic. The shock is assumed to have a Mach number $\mach=3$ at the radius of one Mpc from the center of a typically massive cluster. The SZ effect causes a negative signal boost% roughly
~of roughly the same spatial scale as the radio synchrotron emission. In Figure~\ref{fig:radiosz} (right), the impact of this flux contamination is shown for fiducial relic spectra with different Mach numbers. The relative effect is stronger at lower Mach numbers, where the synchrotron {spectrum is steeper}. {In Figure~\ref{fig:simobs}, we demonstrate that since the SZ signal variation  occurs roughly at the same length scale} as the relic synchrotron emission, interferometric measurements trying to obtain the GHz-frequency relic fluxes will not be immune to SZ flux contamination, even though the global SZ signal from the cluster itself will be almost completely filtered out.

We showed in \cite{SZrelic} that the flux contamination can be {from} a few percent to over 80\% in the frequency range of 10--20 GHz for several well-known cluster relics. For the measured flux values in the Sausage and Toothbrush relics, this simple model provides an adequate fit (Figure~\ref{fig:relicfit}). In the Sausage relic, the 16 GHz data point appears to be low in comparison to the models, but still within the range of a steep spectrum synchrotron emission ($\mach \lesssim 3$). {A similarly low Mach number} for the Sausage relic shock is actually inferred from the X-ray observations \cite{Aka13}. We also note that a steep spectrum is not a necessary condition for the SZ contamination to be of significance; in these examples, it is suggested that way because we are trying to fit the full relic spectrum with a single power-law model. The difficulty of fitting the low-frequency part with a single power-law for these relics is well known, and in reality, the shock fronts can have a distribution of Mach numbers (e.g.,~\cite{Skill13}). Our goal is to show that the SZ flux contamination is {non-negligible and can even be the dominant effect} behind the observed spectral steepening of radio relics at GHz frequencies. This hypothesis can furthermore lead to new methods for testing the relic-shock connection and putting constraints on the relic magnetic field or the shock acceleration efficiency (see \cite{SZrelic}).

\begin{figure}[h]
\centering
\includegraphics[width=9.2cm, clip=true, trim=-20 -40 0 0]{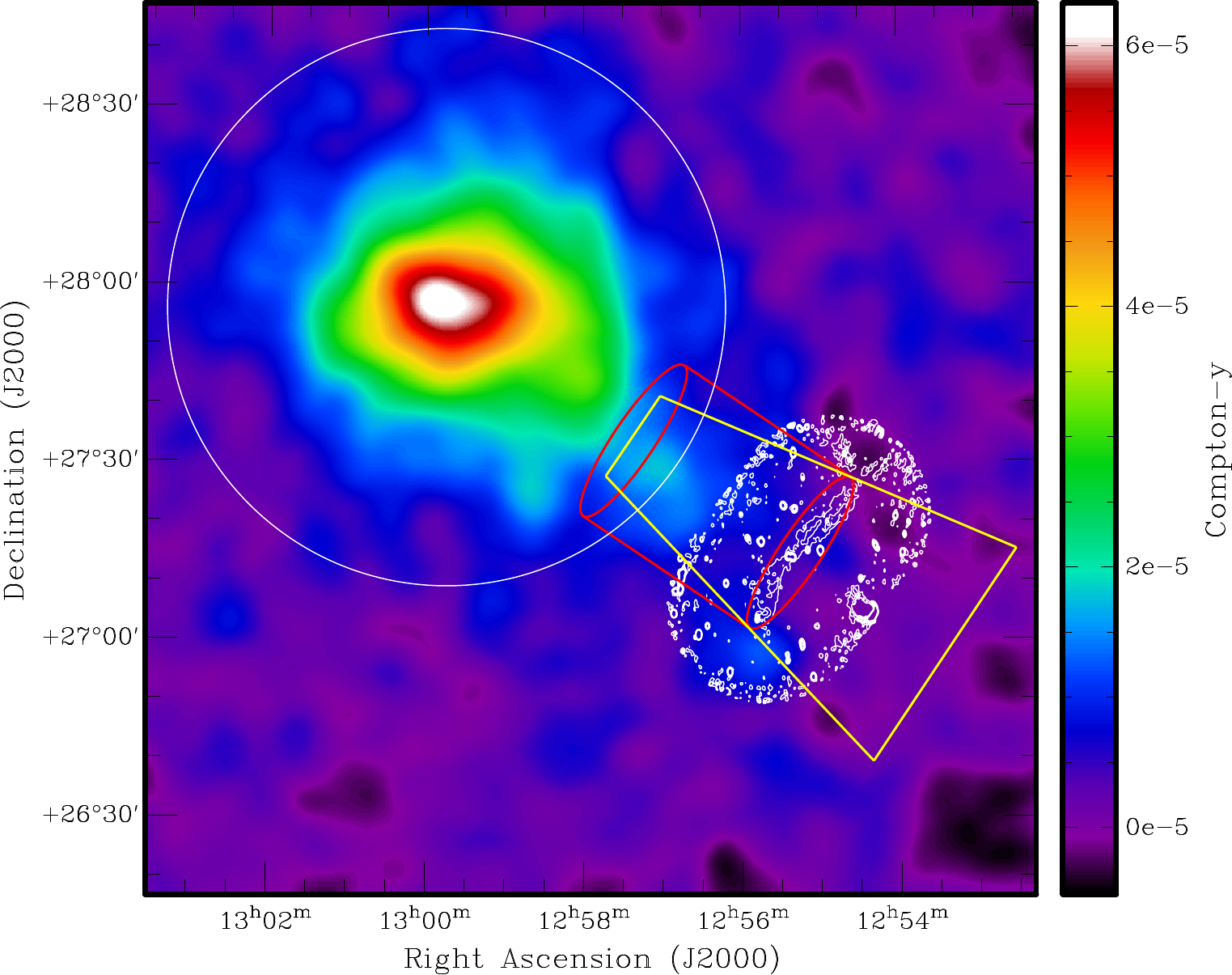}
\includegraphics[width=8.4cm, clip=true, trim=15 0 0 0]{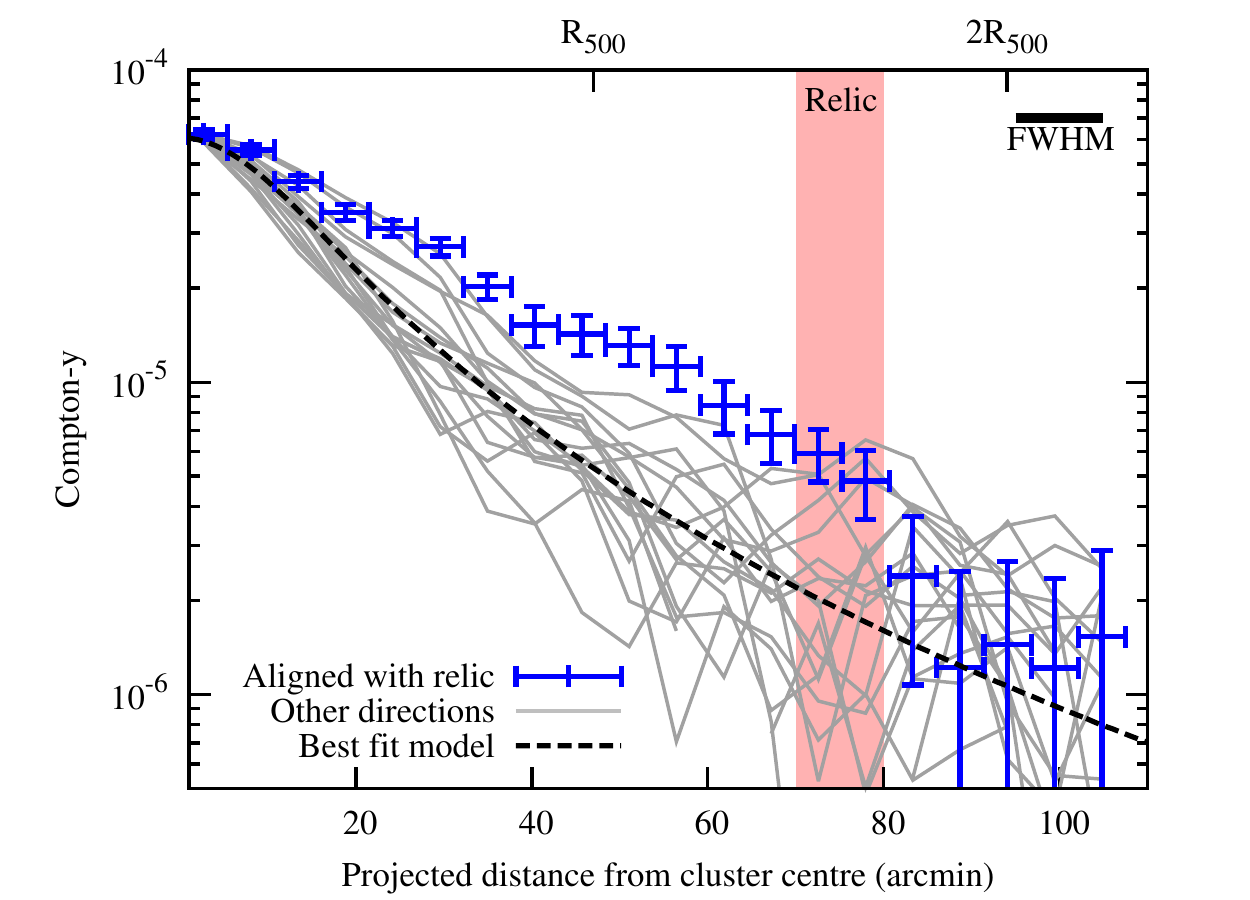}%
\includegraphics[width=9.0cm, clip=true, trim=34 0 -30 -10, height=6cm]{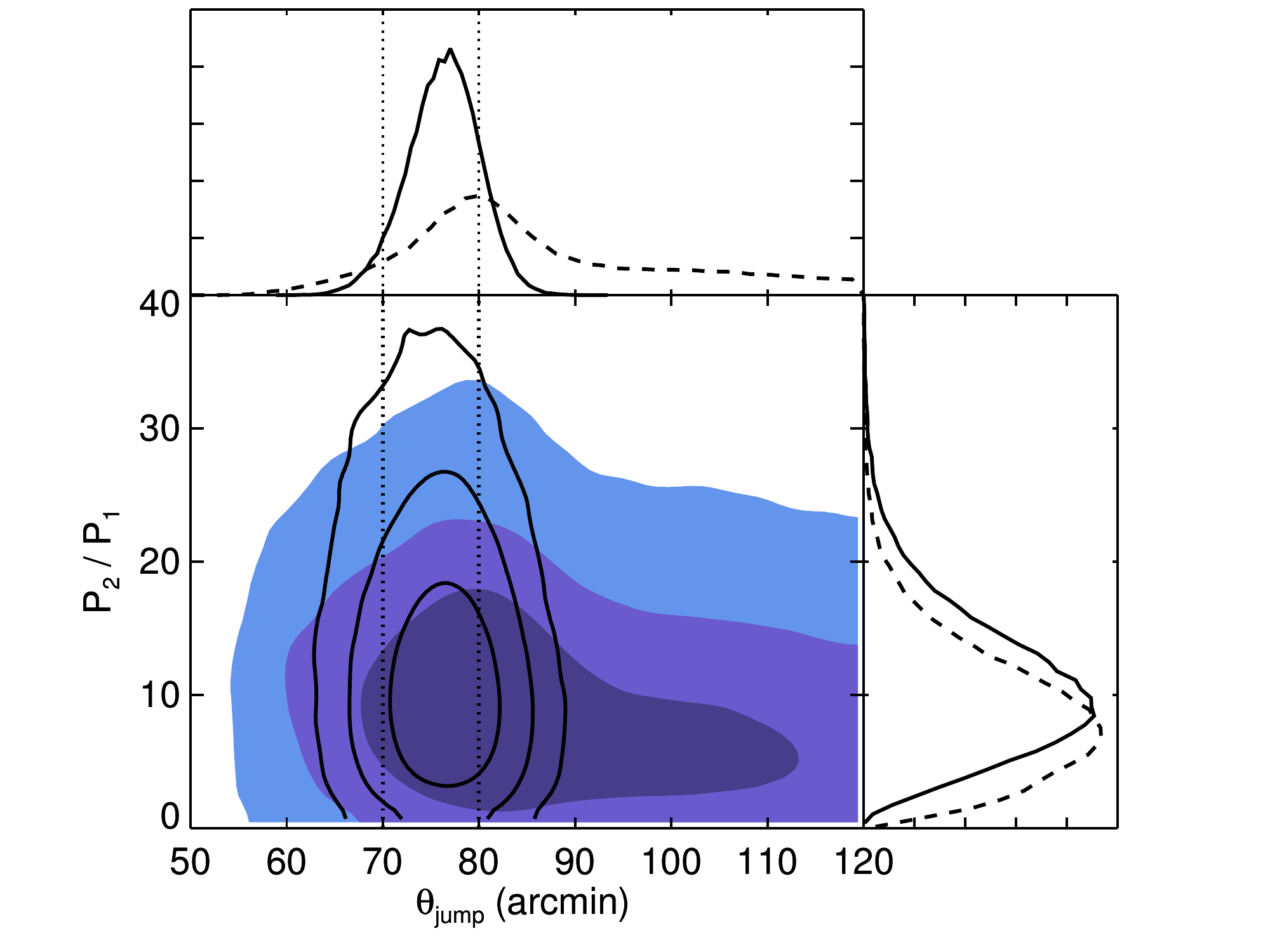}
\caption{The relic shock in the Coma cluster as detected by the SZ effect (from \cite{Er15}). \textbf{Top:} {Coma} Comptonization map derived from a linear combination of the \plk data, overlayed with 2.3 GHz radio contours from the Westerbork Synthesis Radio Telescope (WSRT) array. The thin white circle denotes the $R_{500}$ of Coma, and the conical sector used for profile fitting and the cylindrical {shock model geometry} are shown respectively in yellow and red. 
\textbf{Lower left:} SZ profile in the direction of the relic shock, shown in blue. Various gray lines denote similar profiles in the other directions, and the dashed line is the best-fit model averaging these other sectors (see \cite{Er15} for details). 
\textbf{Lower Right:} Joint likelihood contours for the shock position and the pressure ratio, with (solid lines) and without (dashed lines) having a radio prior on the shock location. FWHM: full width at half maximum.}
\label{fig:comadata}
\end{figure}

%%%%%%%%%%%%%%%%%%%%%%%%%%%%%%%%%%%%%%%%%%
 
\section{The first SZ effect measurement of a relic shock}

We have discussed the potential impact of the SZ effect in measuring radio relic fluxes at GHz-frequencies, but are these hypothesized  pressure boosts from the relic shocks actually observable?
{For answers, we turn to direct SZ measurements of radio relics at 100 GHz or above (at these frequencies, synchrotron contamination to the SZ signal similar to what we discussed in Section \ref{sec:szcont} can be safely ignored).}
Our work {reported on} the first SZ measurement of a relic shock, by {making use of the publicly available \plk data}, at the location of the Coma cluster relic \cite{Er15}. This Coma relic shock had been measured previously in the X-ray, using {\it XMM-Newton} and {\it Suzaku} data \cite{Ogr13,Aka13}, but a debate persisted on whether it is an infall shock caused by accreting materials, or a merger shock originating from the core region of the Coma from a {past} merger \cite{BR11,Ogr13}. Our measurements show clear preference for the merger shock scenario.

The work presented in \cite{Er15} was based on a Comptonization map (or $y$-map) extracted from the 15-month nominal \plk mission data, and we additionally made use of a high-resolution 2.3 GHz radio measurement of the Coma relic to set a prior on the shock location. The \plk SZ image is shown in Figure~\ref{fig:comadata} (top), along with the radio contours for the relic at 2.3 GHz. The SZ profile in the direction of the shock propagation is shown in the lower-left panel, in comparison to profiles from other radial directions. The apparent significance of the $y$-jump is not high; nevertheless, \plk data give clear evidence for a shock (${\cal M} = 2.7^{+0.8}_{-0.9}$ from a cylindrical shock model). 
This result is corroborated when the final (2015) data release from \plk is used; for example, using a spherical geometry as is common in cluster shock modeling, we obtained ${\cal M} = 2.2 \pm 0.3$ \cite{SZrelic}, which is fully consistent with the original results from \cite{Er15}.
%based on the same cylindrical shock model the more recent  data provide a tighter constraint (${\cal M} = 3.4 \pm 0.5$)
Models based on an infall shock or a pressure excess associated with a sub-cluster are ruled out when \plk results are combined with  X-ray measurements. Importantly, \plk data alone show evidence for a pressure jump {in the Coma relic} without any additional radio or X-ray priors (Figure~\ref{fig:comadata} lower-right), {thus making it an independent confirmation of the existence of a shock}.

%%%%%%%%%%%%%%%%%%%%%%%%%%%%%%%%%%%%%%%%%%

\section{ALMA-SZ measurements of a relic shock}

{Using \plk $y$-maps with roughly $10\amin$ resolution is clearly not the optimal method for measuring relic shocks; nevertheless}, it was made possible thanks to the proximity of the Coma cluster. As the X-ray observations evolved from {\it Uhuru} to \chan to provide excellent spectral-imaging measurements of shock fronts in several galaxy clusters, the SZ effect now has a similarly powerful instrument in ALMA, to make arcsecond resolution images of shock fronts in the millimeter/sub-millimeter wavebands. We have recently reported the first measurement of a cluster merger shock with ALMA \cite{ALMA}, which is also one the first ALMA results on galaxy cluster SZ effect {in general}. The target was the highest redshift radio relic currently known, in the famous El Gordo cluster at $z=0.87$ \cite{Menan12,Lind14}, in a clear demonstration of the power of the SZ effect to make resolved cluster images with equal efficiency at all redshifts.

\begin{figure}[t]
%\centering
\hspace{-8mm}
\includegraphics[width=10cm, clip=true, trim=0 -15 0 0]{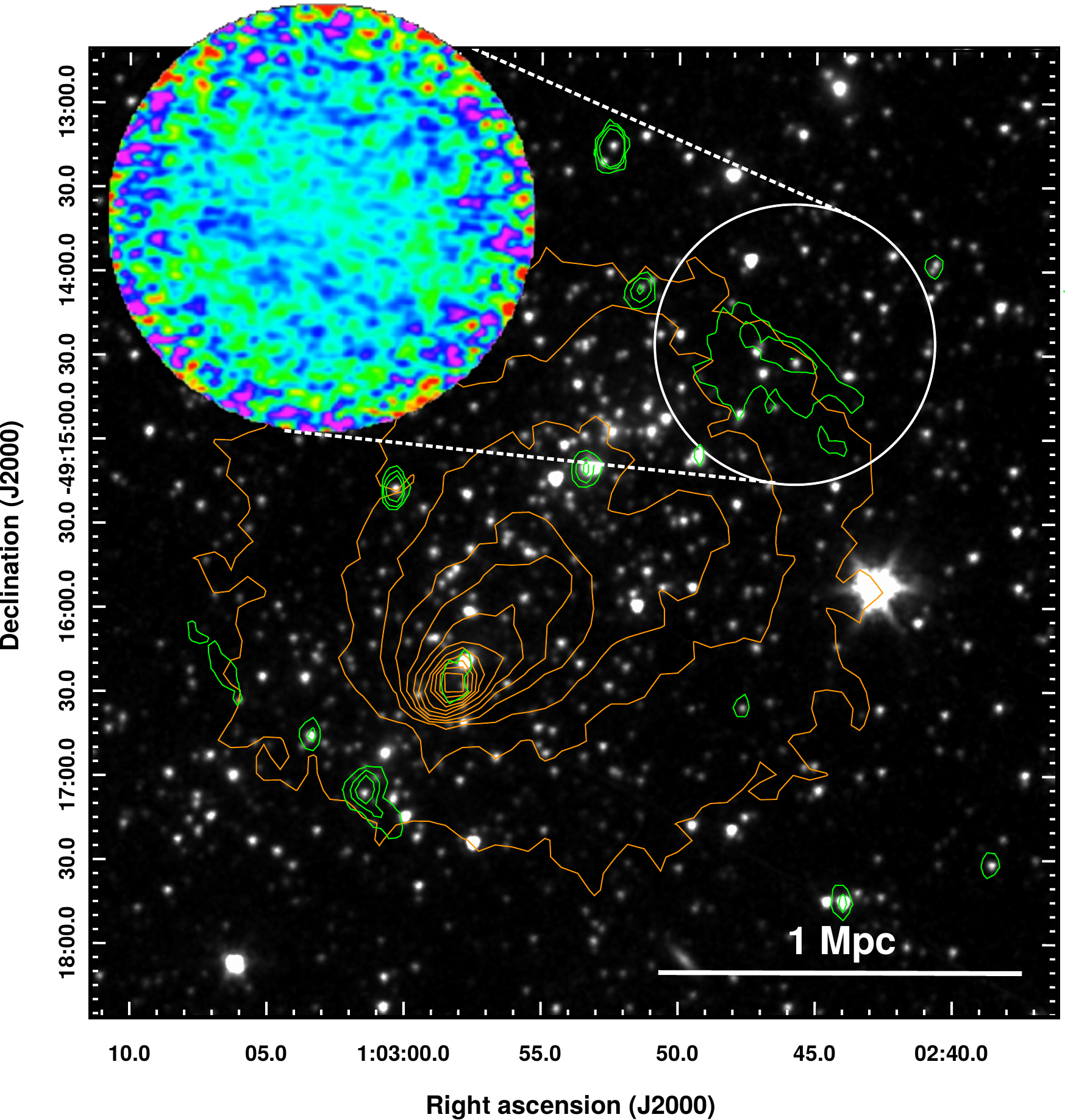}%
\includegraphics[width=7.2cm, clip=true, trim=0 -10 0 0]{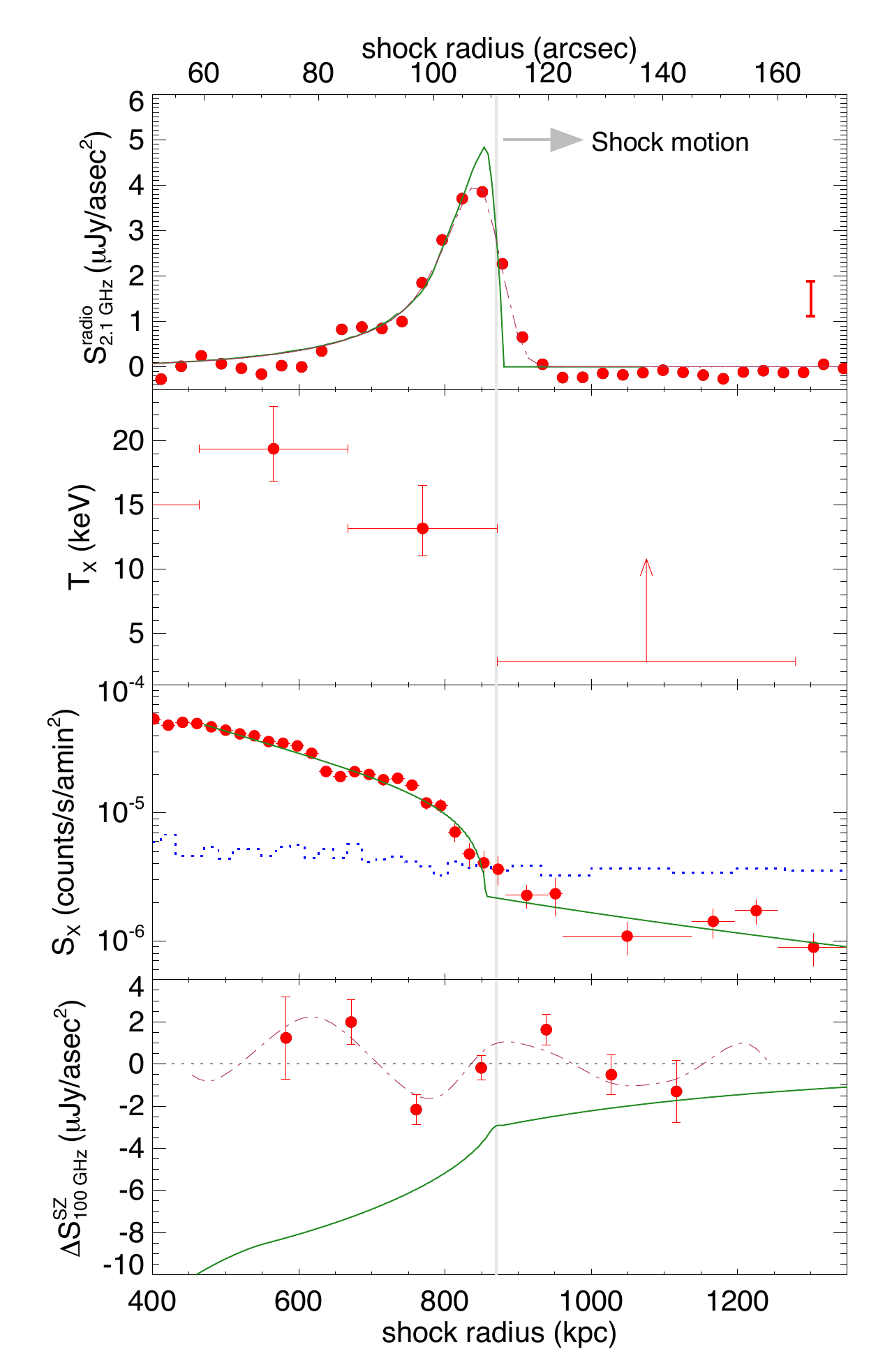}
\caption{A relic shock in the El Gordo cluster at $z=0.87$ (from \cite{ALMA}). \textbf{Left:} Multi-wavelength view of the El Gordo cluster and its NW relic. The background image is a {\it Spitzer} IRAC mosaic at $3.6 ~\mu$m, and the orange and green contours mark the \chan X-ray $0.5-2$ keV emission and the ATCA 2.1 GHz radio emission, respectively. The observed SZ intensity distribution {from an ALMA deconvolved image} is shown in the zoomed-out inset. \textbf{Right:} Thermal and non-thermal signal variations across the relic shock. Shown here from top are the 2.1 GHz radio synchrotron profile, the \chan X-ray temperature measurements, the X-ray brightness profile across the relic, and the ALMA SZ measurement together with the best fit shock model (green lines). }
\label{fig:gordo}
\end{figure}

The small field of view of ALMA at 100 GHz (primary beam FWHM $\sim 1\amin$) allows for the imaging of only a limited sector of a galaxy cluster with a single pointing, but the angular resolution (in a compact array configuration) is {roughly $3\asec$ and unparalleled for SZ observations}. Our observation of the prominent NW relic in the El Gordo cluster using a brief 3 hr ALMA exposure is depicted in Figure~\ref{fig:gordo} (left panel). The background image is a mosaic of {\it Spitzer} IRAC pointings at 3.6 \textmu m, and the colored contours show the soft-band X-ray (from {\it Chandra}) and 2.1 GHz radio synchrotron \linebreak(from ATCA)
~emissions. With ALMA, we were able to clearly detect the signature of an underlying pressure discontinuity from {an observed} modulation of the SZ signal, as shown by the deconvolved image (or dirty image) in the inset. The shock modeling was not based on this image, however, but used a novel MCMC
~fitting technique directly applied to the visibility data (or $uv$-data) in order to avoid imaging biases. {This Bayesian method also enables an easy combination of results from the X-ray and radio wavebands.}

The ALMA SZ data provides evidence for a shock whose amplitude and location is consistent with the radio and X-ray measurements, as shown in Figure~\ref{fig:gordo} (right panel). This is the first instance where the thermal and non-thermal signal variations are modeled self-consistently across a relic shock based on the same underlying shock geometry. The \chan exposure totals roughly 100 hr (350 ks), but even with that deep data, the outer (pre-shock) region's temperature cannot be constrained for this high-$z$ object. SZ measurements can thus be an indispensable tool to robustly constrain the shock Mach numbers from many high-$z$ radio relics, {soon to be discovered from multiple} upcoming radio surveys.
{More broadly, joint SZ/X-ray analysis combining high-resolution SZ observations with short-exposure X-ray data---where the latter would provide constraints on the gas density but not necessarily its temperature---can be an efficient and wide-ranging tool for modeling the shock velocities and physical conditions in the pre-shock medium.}

\begin{figure}
%\centering
\hspace{-6mm}
\includegraphics[width=8cm]{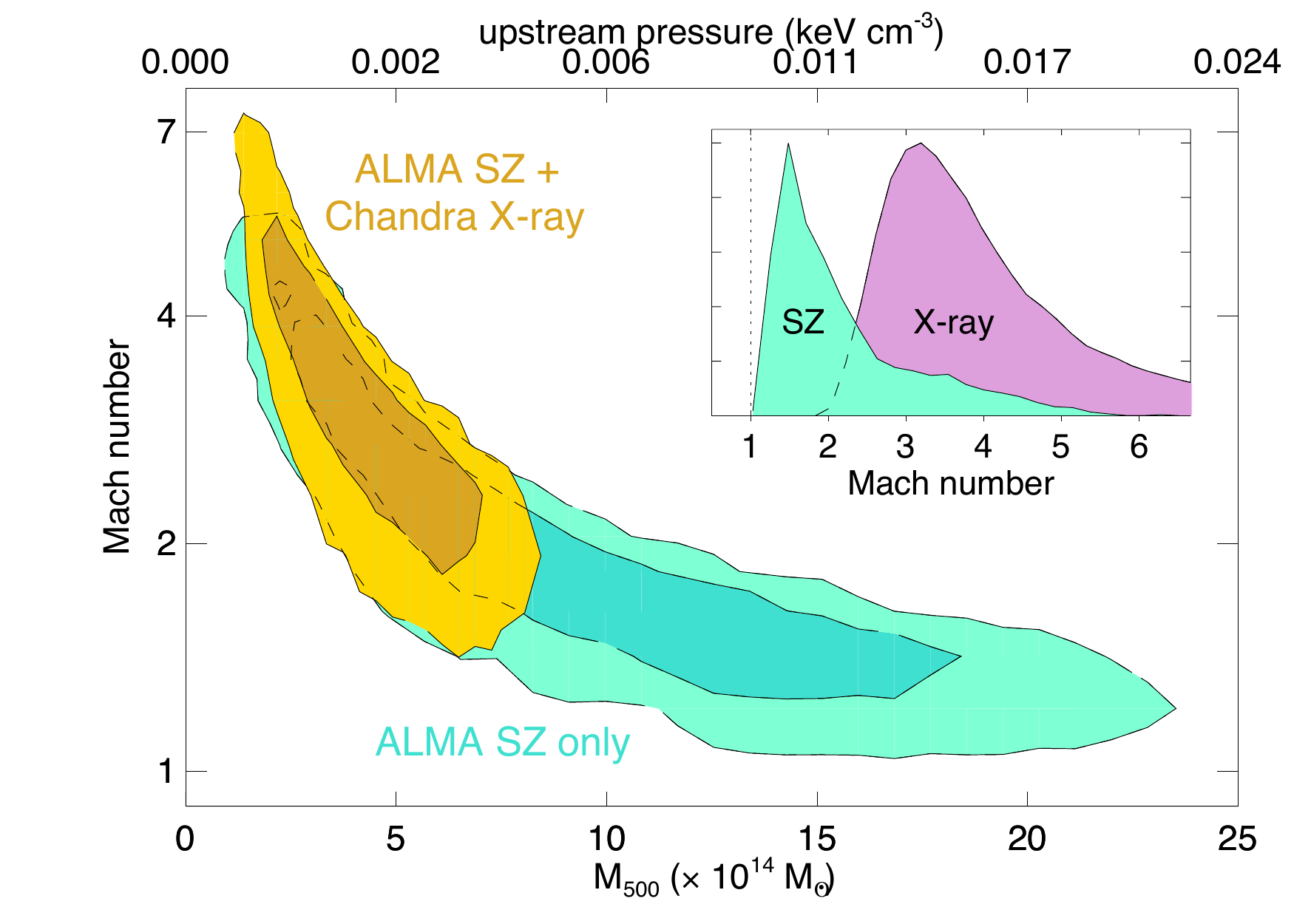}%
\hspace{5mm}
\includegraphics[width=8cm]{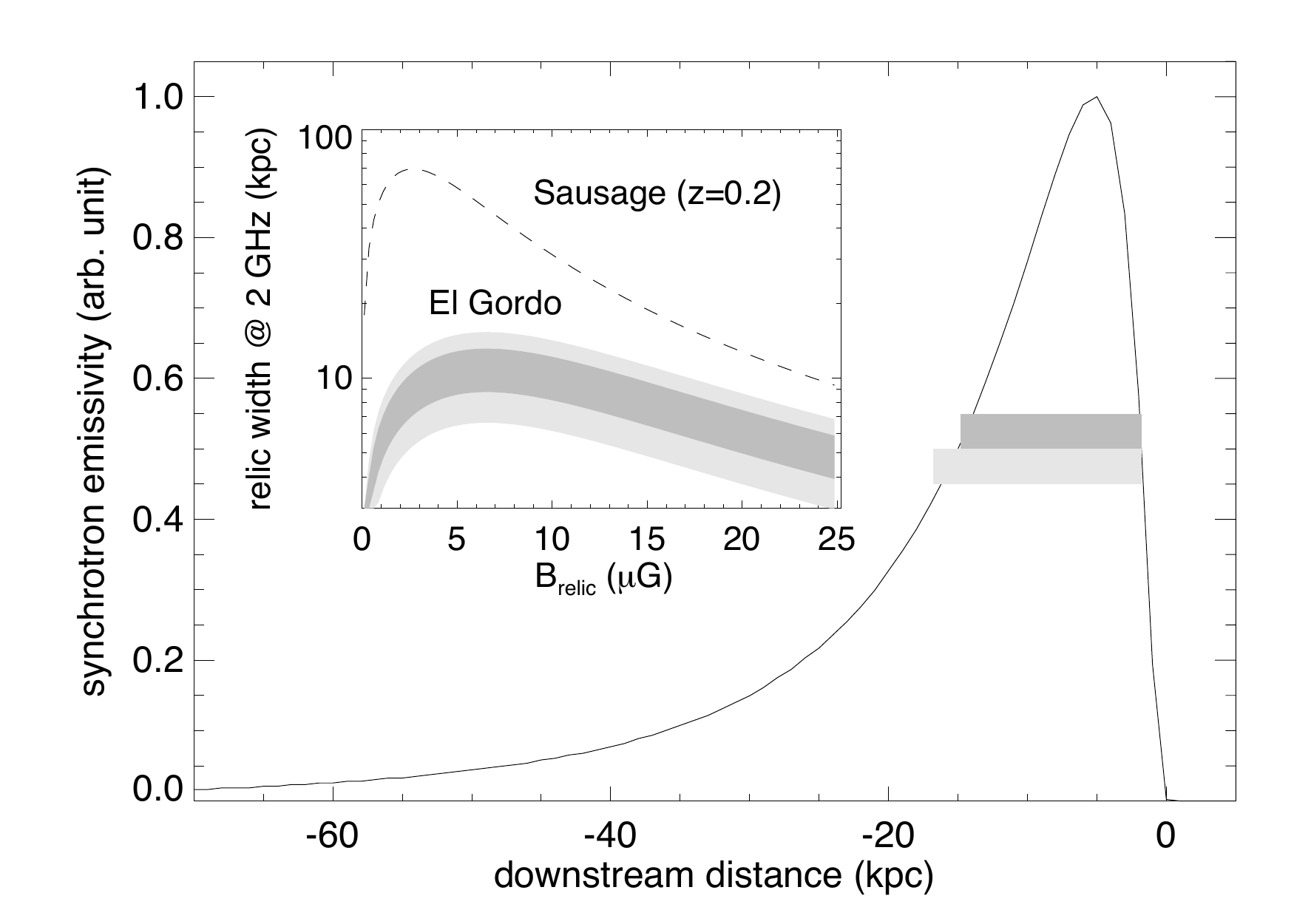}
\caption{%Illustration of science 
Results from the thermal/non-thermal modeling of the El Gordo NW relic shock (from~\cite{ALMA}). \textbf{Left:} Joint likelihood contours for the shock Mach number and the cluster mass/pre-shock pressure. {Darker and lighter colors mark the 68\% and 95\% credible regions.
%where ALMA-only results are shown in green and joint SZ/X-ray results in yellow contours, respectively.
The inset figure shows the marginalized Mach number distribution from independent SZ and X-ray analyses.}
\textbf{Right:} Relic magnetic field strength from the width of its synchrotron emissivity region. The solid line shows a deprojected emissivity model, and the horizontal grey bands mark the maximum relic width at 68\% (dark grey) and 95\% (light grey) confidence, as estimated from a joint modeling of the radio synchrotron and the X-ray/SZ data.}
\label{fig:science}
\end{figure}   

Two {present} examples of joint SZ/X-ray modeling ({also in combination with radio synchrotron data}) are shown in Figure~\ref{fig:science}. In the left panel, we illustrate the Mach number constraints. SZ/X-ray joint fit suggests a similar Mach number value as in a previous radio spectral slope analysis \cite{Lind14}, but there is a mild tension between the individual X-ray and SZ measurements, as shown in the left panel inset.
{ALMA SZ modeling by itself points towards a weak shock, with peak likelihood for the Mach number around $\mach \sim 1.5$, whereas the X-ray brightness modeling supports a strong shock with $\mach \gtrsim 3$.}  
{Although statistically not very significant, we can speculate on some physical effects that can cause a similar difference between the SZ and X-ray Mach number measurements. Such effects can include a 
possible non-equilibrium between the electron and ion temperatures or a boost in the X-ray brightness from an inverse Compton emission behind the shock front. The mismatch can also be partially due to a projection bias, where the choice of the shock geometry affects the X-ray brightness and the Comptonization profiles differently.}

In the right panel of Figure~\ref{fig:science}, the magnetic field strength inside the radio relic is constrained from {a} deprojected synchrotron emissivity profile, making use of an empirical model described in \cite{SZrelic}.  The upstream temperature and the shock Mach number are constrained from the SZ/X-ray analysis when calculating the relic width as a function of the magnetic field strength (results shown in the inset of the right panel). The estimated field {strength} is of the order 4--10 \textmu G, which is unusually high for this redshift. It is possible that part of the observed relic width at 2 GHz comes from a secondary amplification of the particle energies or the magnetic fields \cite{Fuji15,Don16}.

%%%%%%%%%%%%%%%%%%%%%%%%%%%%%%%%%%%%%%%%%%
\section{Conclusions}

We presented the first SZ measurements of galaxy cluster merger shocks at the location of radio relics. {These are also} the first SZ-measured shocks in the cluster outskirts. Even though SZ measurements alone cannot provide a complete thermodynamical description of the ICM \linebreak(i.e., independent density and temperature measurements) as in X-ray imaging/spectral analyses, SZ imaging is clearly adequate for modeling the location, strength, and the geometry of a merger shock. Especially in the cluster outskirts and for high-$z$ objects, SZ measurements should be the preferred tool, as it scales linearly with the gas density and its surface brightness is redshift independent.
{For a complete thermodynamical description of the pre-shock medium and an estimation of the shock velocities, joint SZ/X-ray analysis combining short-exposure X-ray and SZ data can be far more efficient than deep X-ray observations alone.} 
{For radio observers,} we have shown that the SZ effect can act as an effective contaminant to the synchrotron flux measurements at GHz frequencies (1--30 GHz), lowering the measured fluxes and giving the impression of a steepening synchrotron spectrum. However, far from being a simple nuisance for relic measurements, this SZ flux modification can potentially offer some of the best evidence for the relic-shock connection, and {under certain assumptions can} provide an independent method for estimating the relic magnetic field or the shock acceleration efficiency.

\bigskip
%%%%%%%%%%%%%%%%%%%%%%%%%%%%%%%%%%%%%%%%%%
\acknowledgments{We acknowledge the help from our various collaborators in the original works and the numerous colleagues for discussions. JE, MS, and FV have been supported by the grants SFB 956, TR 33, and  VA 876/3 from the Deutsche Forschungsgemeinschaft (DFG), respectively. JE also acknowledges support through the IMPRS and BCGS of Cologne/Bonn. This article uses results made from the following ALMA data: ADS/JAO.ALMA\#2015.1.01187.S . ALMA is a partnership of ESO (representing its member states), NSF (USA) and NINS (Japan), together with NRC (Canada), NSC and ASIAA (Taiwan), and KASI (Republic of Korea), in cooperation with the Republic of Chile. The Joint ALMA Observatory is operated by ESO, AUI/NRAO and NAOJ. In addition, we make use of public data taken by the \chan X-ray Observatory (CXO), the Australia Telescope Compact Array (ATCA), the Westerbork Synthesis Radio Telescope (WSRT), the Effelsberg  100m radio telescope, and the \plk surveyor. Figures 1, 2, and 3 are partially reproduced with permission from \copyright ~ESO.}

%%%%%%%%%%%%%%%%%%%%%%%%%%%%%%%%%%%%%%%%%%
\authorcontributions{KB initiated the work on the cluster radio relic SZ measurements and lead-authored two papers; JE was the lead author of the Coma relic paper and analyzed {\it Planck} data; MS contributed to the ALMA data analysis and mock interferometric observations; FV helped with the theoretical modeling of radio relic shocks; and DE helped to analyze the {\it Chandra} X-ray data for the El Gordo relic.}

%%%%%%%%%%%%%%%%%%%%%%%%%%%%%%%%%%%%%%%%%%
\conflictofinterests{The authors declare no conflict of interest.}

\bibliographystyle{mdpi}

\renewcommand\bibname{References}

%=====================================
% References, variant B: external bibliography
%=====================================
%\bibliography{ewass_references}

%%%%%%%%%%%%%%%%%%%%%%%%%%%%%%%%%%%%%%%%%%
%% optional

%%%%%%%%%%%%%%%%%%%%%%%%%%%%%%%%%%%%%%%%%%
\end{document}